\documentclass{article}

\usepackage{arxiv}

\usepackage[utf8]{inputenc} 
\usepackage[T1]{fontenc}    
\usepackage{hyperref}       
\usepackage{url}            
\usepackage{booktabs}       
\usepackage{amsfonts}       
\usepackage{nicefrac}       
\usepackage{microtype}      
\usepackage{lipsum}		
\usepackage{graphicx}
\usepackage{natbib}
\usepackage{doi}

\usepackage{float}
\usepackage{booktabs} 
\usepackage{amsmath,amssymb}

\usepackage{authblk}

\title{From Unsupervised to Semi-supervised Anomaly Detection Methods for HRRP Targets}

\date{} 					

\author[1, 2]{\small Martin Bauw}
\author[1]{\small Santiago Velasco-Forero}
\author[1]{\small Jesus Angulo}
\author[2]{\small Claude Adnet}
\author[2]{\small Olivier Airiau}

\affil[1]{\footnotesize Center for Mathematical Morphology, MINES ParisTech, PSL Research University, France}
\affil[2]{\footnotesize Thales LAS France, Advanced Radar Concepts, Limours, France}

\renewcommand{\headeright}{HAL preprint}
\renewcommand{\undertitle}{HAL preprint}
\renewcommand{\shorttitle}{From Unsupervised to Semi-supervised AD Methods for HRRP Targets}

\hypersetup{
pdftitle={A template for the arxiv style},
pdfsubject={q-bio.NC, q-bio.QM},
pdfauthor={David S.~Hippocampus, Elias D.~Striatum},
pdfkeywords={First keyword, Second keyword, More},
}

\begin{document}
\maketitle

\begin{abstract}
Responding to the challenge of detecting unusual radar targets in a well identified environment, innovative anomaly and novelty detection methods keep emerging in the literature. This work aims at presenting a benchmark gathering common and recently introduced unsupervised anomaly detection (AD) methods, the results being generated using high-resolution range profiles. A semi-supervised AD (SAD) is considered to demonstrate the added value of having a few labeled anomalies to improve performances. Experiments were conducted with and without pollution of the training set with anomalous samples in order to be as close as possible to real operational contexts. The common AD methods composing our baseline will be One-Class Support Vector Machines (OC-SVM), Isolation Forest (IF), Local Outlier Factor (LOF) and a Convolutional Autoencoder (CAE). The more innovative AD methods put forward by this work are Deep Support Vector Data Description (Deep SVDD) and Random Projection Depth (RPD), belonging respectively to deep and shallow AD. The semi-supervised adaptation of Deep SVDD constitutes our SAD method. HRRP data was generated by a coastal surveillance radar, our results thus suggest that AD can contribute to enhance maritime and coastal situation awareness.
\end{abstract}


\section{Introduction}

High resolution range profiles (HRRP) targets identification using machine learning methods is an active field of research, thanks to the progress of machine learning research and its arrival in the field of radar signal processing. In particular, neural networks demonstrated promising results for HRRP target recognition \cite{feng2017radar} \cite{zhao2018radar}. Alongside supervised learning approaches which allow multi-class classification pipelines, unsupervised and semi-supervised anomaly detection (AD) benefit from the recent advances in machine learning and are part of the solutions to the current issues encountered in radar targets identification. Unsupervised and semi-supervised AD (SAD) allow automatic detection of targets for which no or very few data samples are available. In order to achieve such detection, the targets should be kept outside the definition of a normal environment, possibly including normal targets, which will on the other hand be described by a significant majority of data points. Once normality is defined, one-class classifiers, or outlier and novelty detection methods will be able to single out targets by detecting a deviation from a normal situation in the watched environment. 

AD is already used in various tasks outside of HRRP target recognition within the radar community: \cite{laxhammar2008anomaly} for example used Gaussian Mixture Model as a cluster model to achieve AD on velocities and positions. Going back to HRRPs processing, Auto-encoders (AE) were used in \cite{wagner} and \cite{wan2019convolutional} respectively to detect targets in background profiles and to reject outliers using a weighted training objective combining HRRP rejection and recognition. In these two cases, AEs are used for AD thanks to their reconstruction error, assumed to be high for anomalous samples when trained on normal ones, as was done in \cite{sakurada}. A one-class classifier was also used on HRRP data in \cite{du2019hrrp} to reject clutter generated false alarms.

This work presents a use case of AD methods with 1D radar range profiles, and intentionally gathers results generated with several shallow and deep learning methods, baseline and new, to offer a wide but non-exhaustive overview of AD. The results presented are part of a more extensive effort to explore the full potential of AD to achieve improved radar detection, whether in an unsupervised or semi-supervised setup. The rest of the paper is organized as follows: in section \ref{Anomaly detection methods considered} we describe the unsupervised and semi-supervised AD methods included in our benchmark, in \ref{HRRP dataset} we introduce the nature of the samples in our dataset and offer an operational context justifying our approach. Finally, in \ref{Results} we present details of our experimental setup and discuss the results of the selected unsupervised and semi-supervised AD methods on our HRRP data.

\section{Anomaly detection methods considered}
\label{Anomaly detection methods considered}

This work considers both unsupervised and semi-supervised AD. Since these very notions are not necessarily defined in a unique fashion in the literature as \cite{deepSAD} still recently emphasized, let us define what is understood as unsupervised and semi-supervised in this work. \emph{Unsupervised AD} refers to all experiments in which the training data completely consists of normal samples, \emph{i.e.} samples belonging to the class chosen as normal. Experiments where the training set is polluted by unlabeled anomalies are also considered as unsupervised AD, since such pollution aims at experimenting with flawed training samples collection in the interest of realism, with no adaptation of the one-class classification or outlier detection algorithm. On the other hand, \emph{Semi-supervised AD} refers to experiments where, in addition to the training set samples belonging only to the normal class, a few labeled anomalies are available and taken into account by the training objective to help improve the detection. Intuitively, the labeled anomalies can refine the decision boundary in their surroundings within the latent space used by the SAD.

This study did not consider experiments combining unlabeled anomalies pollution of the training set and the availability of labeled anomalies to mark out normality during training. Furthermore, pollution of training data with unlabeled anomalies and labeled anomalies for unsupervised AD and SAD respectively will always be done using a single pollution anomalous class at a time. Since normality is defined by a single class out of four in our experiments, this means that there will be three unseen classes of anomalies in testing for unsupervised AD without training set pollution, and two unseen classes of anomalies in testing for unsupervised AD with training set pollution and SAD. Evaluating AD with unknown classes of anomalies is critical to respect the unforeseeable diversity of anomalies to be detected. AD indeed translates into a form of cluster assumption regarding the normal class which cannot hold for unspecified anomalies. Further discussion and an information-theoretic overview of anomaly detection can be found in \cite{deepSAD}.

\subsection{Unsupervised anomaly detection}

\subsubsection{Shallow unsupervised AD}

The common shallow AD methods chosen are one-class Support Vector Machine (OC-SVM)\cite{ocsvm}, Isolation Forest (IF)\cite{isolationforest} and Local Outlier Factor (LOF)\cite{breunig2000lof}. The first method is an extension to one-class classification of the now classic SVM classifiers. OC-SVM projects data in a feature space where it will try to find a maximum margin hyperplane to separate data points from the feature space origin. This is achieved thanks to the following objective function:
\begin{equation}
\min_{w, \rho, \xi} \hspace{0.2in} \frac{1}{2} \lvert\lvert w \lvert\lvert^2_{F} - \rho + \frac{1}{\nu n} \sum^n_{i=1} \xi_i
\end{equation}
$\rho$ is the distance separating the origin from the hyperplane $w$, $\xi_i$ are slack variables allowing boundary violation with penalization. $\vert\vert w \vert\vert_F$ regularizes the definition of the hyperplane $w$ using the norm of the feature space $F$ in which data points are projected by a kernel. The integer $n$ is the number of data samples available for training and $\nu \in (0,1]$ is an upper bound on the fraction of outliers during training and a lower bound on the fraction of support vectors for the hyperplane boundary \cite{ocsvm}. IF uses recursive partitioning on subsets of data points in the feature space, and produces an anomaly score based on the ease with which each point is isolated from the rest in each subset. It works based on the assumption that anomalous samples are more susceptible to isolation in the feature space. One recursive partitioning isolation binary tree is built for each subset of data points, making IF an ensemble method. In the end, IF computes an anomaly score $s$ for each instance $x$ whose expression is:
\begin{equation}
s(x,n) = 2^{-\frac{E(h(x))}{c(n)}}
\end{equation}
$n$ is, as for OC-SVM, the number of data points available for training and $c(n)$ the average path length of an isolation tree. The average path length intuitively translates into the average number of recursive splits needed to isolate a data point in the feature space. The vector $x$ is the sample whose anomaly score we want to obtain, and $h(x)$ the associated path length. The previous equation uses $E(h(x))$, the average path length across the forest of isolation trees, normalized by $c(n)$, to obtain $s(x,n)$. IF is a particularly interesting shallow AD method since it is advertised as being able to provide good performances with a small subsample of data and few isolation trees.

The last common shallow AD method considered is LOF. To compute LOF, we choose a certain number $k$ of nearest neighbors to be considered for each data point. The local density attached to a point will be determined by how close its $k$ nearest neighbors are. A point having a higher local density than its neighbors will be more likely to be an inlier, since this translates into belonging to a higher density part of the feature space. Thus, LOF assigns to each data point an outlier score based on the ratios of its own local density and the local densities of its $k$ nearest neighbors.

The innovative shallow AD method used in our benchmark, called Random Projection Depth (RPD), is based on \cite{rpd} \cite{rpd2} and orders data samples from common to outlier values using a statistical depth determined by random projections uniformly distributed in the \textit{d}-dimensional hypersphere. The anomaly score obtained for each sample is actually a stochastic approximation of the RPD by means of a finite set of random projections. The approximation of the projection depth of a data point $x$ belonging to a set of points described by the matrix $X=[x_1, ..., x_n]$ using $p$ random projections is:
\begin{equation}
RPD(x;p, X) = \dfrac{1}{1 + O(x;p,X)}
\end{equation}
with $O(x;p,X)$ an outlyingness \cite{rpd2} defined by
\begin{equation}
O(x;p, X) = \max_{u \in \mathbb{U}} \dfrac{\vert u^Tx - MED(u^TX) \vert}{MAD(u^TX)}
\end{equation}
where $\mathbb{U} = \{u_1, ..., u_p\}$ are $p$ vectors belonging to the hypersphere $\mathbb{S}^{d-1}=\{x \in \mathbb{R}^d:\vert\vert x \vert\vert_2 = 1\}$. In the previous equation $MED$ is the median and $MAD$ the median absolute deviation. One should note that the number of projections $p$, a parameter for our AD implementation, needs to be big enough to produce a relevant statistical depth approximation.

\subsubsection{Deep unsupervised AD}

A single common deep baseline was chosen for this study, the AD autoencoder\cite{sakurada}. To be more specific, we chose a convolutional autoencoder similar to the one used in \cite{deepSVDD}. This architecture benefits from useful invariance properties. The autoencoder directly uses the reconstruction error $\vert \vert x - \hat{x} \vert \vert^2$ of the generative neural network as an objective function and anomaly score: being trained only with normal samples, the neural network is assumed to be less able to generate good reproductions of anomalous inputs.

The innovative deep unsupervised AD method of this study is Deep Support Vector Data Description (Deep SVDD) \cite{deepSVDD}. Deep SVDD projects data points within a normality hypersphere using a neural network. This neural network is trained using an objective function penalizing the distance between the normality hypersphere center and the training data points projections, defined by the following expression:
\begin{equation}
\label{deepSVDD}
\min_{W} \dfrac{1}{n} \sum_{i=1}^n \vert \vert \Phi(x_i;W) - c \vert \vert^2+\dfrac{\lambda}{2} \sum_{l=1}^L \vert \vert W^l \vert \vert^2_{Fr}
\end{equation}
$c$ is the hypersphere center, $W$ the neural network parameters, $\vert \vert . \vert \vert_{Fr}$ the Frobenius norm, $x_i$ the $i$-th training sample. The second term is a regularization term penalizing high values in the network parameters across its $L$ layers, with an hyperparameter $\lambda$ to choose the balance with the main training objective. The expression $\Phi(x_i;W)$ denotes the projection of the $i$-th training point in the normality hypersphere latent space using the neural network of parameters $W$. Once trained, the anomaly score $s(x_t)$ of a test sample $x_t$ is $s(x_t)= \vert \vert \Phi(x_t;W^*) - c \vert \vert^2$, where $W^*$ are the neural network learned parameters. One should note that the hypersphere center $c$ is defined by the mean projection of data points after a first pass of the latter in the Deep SVDD network before training. We will therefore compare Deep SVDD initialized with weights from a pretraining network (Deep SVDD pretr on Figure \ref{fig_auc_unsupervised}) with a non pretrained Deep SVDD (Deep SVDD on Figure \ref{fig_auc_unsupervised}) to verify how such initialization can influence the method performances, including through its modification of the hypercenter $c$. This objective function implies all training data points describe normal samples. The intuition behind this neural network is similar to using the encoder of an autoencoder to project data points within a hypersphere in the encoding latent space. OC-SVM, already described in this paper and present in our results, has a close relation with the shallow sister method of Deep SVDD, called support vector data description (SVDD) \cite{svdd}, as pointed out in \cite{deepSVDD}. OC-SVM and SVDD are actually equivalent when using a Gaussian kernel and give similar solutions. Since we use the standard Gaussian (radial basis function) kernel in our experiments with OC-SVM, there was no point in adding SVDD to this study eventhough its name suggests a closer relation with one of our highlighted AD methods. Additionally, as mentioned in \cite{wang2019gods}, let us remind that SVDD makes a strong isotropy hypothesis regarding the latent distribution of data.

\subsection{Semi-supervised anomaly detection}

We present a single SAD method, which is an adaptation of Deep SVDD proposed in \cite{deepSAD} and \cite{deepSADworkshop} to the inclusion of labeled anomalies during training to improve AD performances. SAD is an ideal AD implementation because the SAD paradigm makes use of all data possibly available to obtain optimal detection: it takes into account realistic rare and non representative example anomalies in addition to normal samples to train a model. The inclusion of labeled anomalies in the already described objective function of Deep SVDD is allowed by an additional term in equation \eqref{deepSVDD}. Training this Deep SAD using $n$ unlabeled samples $x_i$ taken as normal and $m$ labeled samples $\widetilde{x}_j$ is performed using the objective:
\begin{equation}
\begin{split}
\min_{W} \dfrac{1}{n+m} \sum_{i=1}^n \vert \vert \Phi(x_i;W) - c \vert \vert^2 \\ + \dfrac{\eta}{n+m} \sum_{j=1}^m (\vert \vert \Phi(\widetilde{x_j};W)-c\vert \vert^2)^{\widetilde{y_j}} +\dfrac{\lambda}{2} \sum_{l=1}^L \vert \vert W^l \vert \vert^2_{Fr}
\end{split}
\end{equation}
$\widetilde{y_j}$ is the label of $\widetilde{x}_j$ of value $-1$ for anomalous samples. Compared to the unsupervised Deep SVDD objective in equation \eqref{deepSVDD}, there is an additional term weighted by the hyperparameter $\eta$ that penalizes labeled training points $\widetilde{x_j}$ when they are near the normality hypersphere center $c$ and anomalous.

\section{HRRP dataset}
\label{HRRP dataset}

\begin{figure}[!t]
\centering
\includegraphics[scale=0.25]{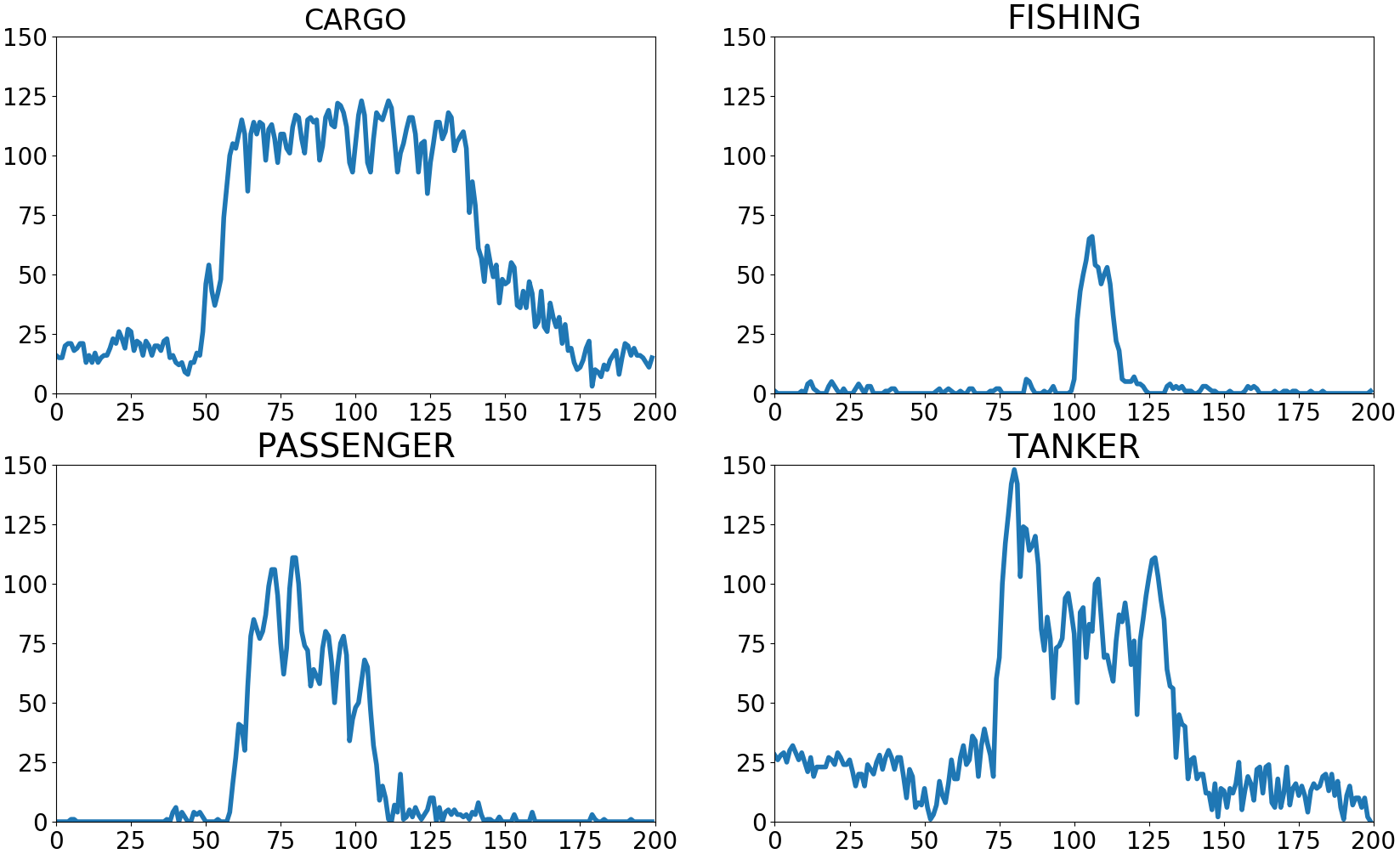}
\caption{Example range profiles (one per class considered). Horizontal axis denotes range cell index, vertical axis denotes amplitude.}
\label{fig_range_profiles}
\end{figure}

The dataset is composed of real radar 1D range profiles (RP) generated by a very high performance radar for coastal surveillance, the Coast Watcher 100 Thales radar. Each range profile is composed of 200 cells with various intensities. The labelling of the data samples stems from the AIS\footnote{Automatic Identification System} announced ship types. This constitutes a first potential difficulty in the creation of the dataset since the labels provided by AIS are functional (\emph{i.e.} they describe the function of a ship, not its appearance and dimensions). Another possible interpretation of our training set pollution with unlabeled anomalies for unsupervised AD is the mislabeling of AIS data, or the intra-class variance of AIS labels, both interpretations affirming the point of experimenting with such pollution. The dataset is balanced and composed of four classes of range profiles: the latter either belong to cargo ships, fishing ships, passenger ships or tankers. A characteristic example RP of each class can be seen on Figure \ref{fig_range_profiles}. A total of 12000 HRRPs are available for each class.

\begin{figure}[!t]
\centering
\includegraphics[scale=0.20]{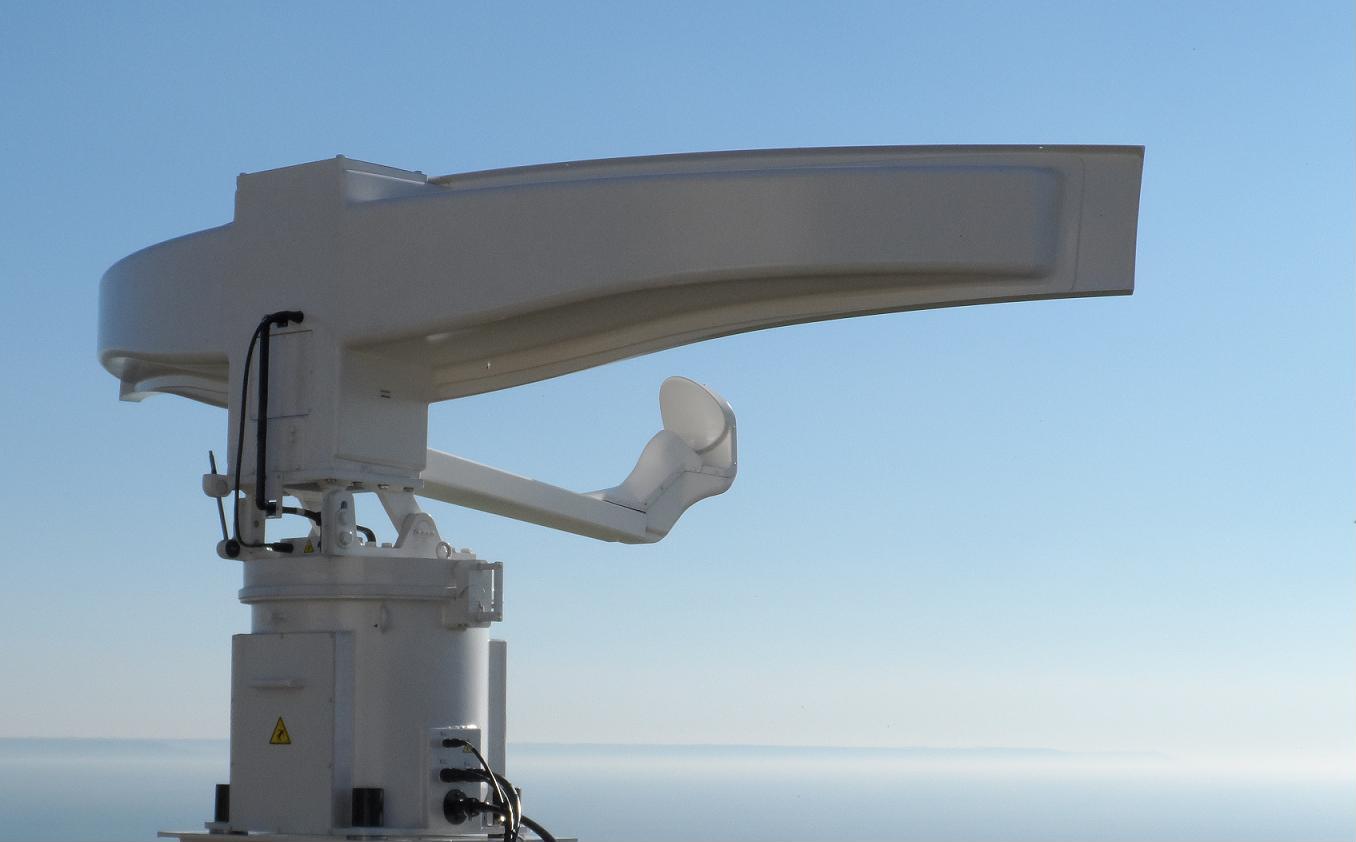}
\caption{Thales Coast Watcher 100 radar, which generated the HRRPs used in this work.}
\label{fig_results}
\end{figure}

\subsection{Operational motivation}

A possible operational use of anomaly detection with the previous classes could be the following: considering a busy but regulated waterway in terms of fishing, an operator could be helped by automatic AD alerts detecting fishing ships in the area. This implies normality is defined by the three other classes (cargo ships, passenger ships, tankers), for which many samples should be available. Note that our experiments define normality based on a single class of ships, but the described operational use can still be achieved by combining the results of AD on each normal class. This scenario can be easily extended to a variety of realistic surveillance contexts: detection of specific ships despite the shutdown of AIS, detection of military ships without AIS and IFF\footnote{Identification friend or foe}.

\subsection{Data preselection}

In order to obtain range profiles expressive enough for relevant experiments, the data samples chosen for our experiments were beforehand selected according to their combined cells energy. This selection aims at avoiding both too small and too high energies as a crude effort to avoid model bias imputable to the outliers of each class. This work deliberately avoids elaborate preprocessing to reveal the potential of AD on raw HRRPs. We will see in our results AD methods applied on data with and without normalisation. Outside this samples preselection and non systematic normalisation, no steps are taken in order to counter amplitude-scale, time-shift and target-aspect sensitivities, which seems uncommon when compared with other approaches \cite{feng2017radar} \cite{du2019hrrp} \cite{wan2019convolutional}. Regarding time-shift sensitivity, one should note that most of our samples are nonetheless approximately aligned. About the diversity of targets, no exact count was made of the ships eventually retained by our preselection, but this diversity is above the one observed in other studies dedicated to HRRP targets such as \cite{wan2019convolutional}.

\section{Results}
\label{Results}

Our unsupervised AD results aim at comparing between shallow and deep AD methods, but also at individually appreciating the sensitivity to training pollution of each unsupervised AD method, with fixed hyperparameters through the pollution changes. We maintain constant hyperparameters when polluting the training set in order to stay relevant regarding an actual implementation of one of the methods considered with imperfect radar data. The results of Deep SAD are here to demonstrate the ability of labeled anomalies to contribute to Deep AD through recent semi-supervised approaches adaptable to HRRP data. Each unsupervised and semi-supervised experiment includes detecting anomalies of at least two classes completely unseen during training at test stage. As was already mentioned, this is fundamental in order to respect the diversity and unpredictability of anomalies. The metric defining all results is the area under the receiver operating characteristic curve (ROC AUC), as was used in other recent works on AD such as \cite{deepSVDD} \cite{du2019hrrp} \cite{wan2019convolutional} \cite{wang2019gods}. For each experiment, over the 48000 HRRPs of our dataset a random fraction of 10\% defines the test set.

\subsection{Experimental setup}

As was done in \cite{deepSVDD}, we use a LeNet-type CNN with leaky ReLU activations for our Deep SVDD and Deep SAD neural network. The CAE architecture will be similar to allow the use of the trained weights of its encoder for the initialization of the Deep SVDD network (such initialization constitutes a pretraining). For these three neural network based AD, a batch size of 128 and a learning rate of $10^{-3}$ were used. The weight decay hyperparameter was set to $10^{-6}$. The CAE used for AD and to provide Deep SVDD with pretrained parameters has been trained during ten epochs. For unsupervised AD, Deep SVDD was trained during 20 epochs whether it was initialized with the CAE encoder parameters or not. For SAD, the Deep SAD network adapted from Deep SVDD with a new objective was trained during 20 epochs, with a semi-supervised objective term balanced by $\eta$ equal to one, as found in \cite{deepSAD}.

Regarding the hyperparameters of our shallow methods: LOF was set with a number of nearest neighbors of 48 to be considered in local densities estimations, and a contamination of 10\%. RPD involves 1000 random projections to produce its statistical depth approximation, and OC-SVM had its $\nu$ set to $10^{-1}$. Our IF was executed with 100 estimators, a contamination of 10\% and a maximum number of samples per subsample of 1024, thus respecting the original spirit of IF, designed to work best with a substantially limited subsample \cite{isolationforest}. The methods parameters were directly inspired by the available implementation of \cite{deepSVDD}.

In the case of dimensionality reduction by PCA, the PCA is systematically preceded by min-max normalization and the number of components kept is chosen in order to retain 95\% of the variance. This dimensionality reduction setup was inspired by the methodology proposed in \cite{deepSVDD}. It is to be noted that the PCA is always fitted using the training samples of the normal class only, this means that for each normal class change, there is a different PCA that is being used. The PCA used for dimensionality reduction of the test samples of a given class will, for instance, not be the same between an experiment where the same class is defined as normal, and another experiment for which another class will define normality and be responsible for the AD training. For Deep AD experiments, only a similar min-max normalization will be applied to the data.

\subsection{Unsupervised AD with training pollution}

\begin{figure*}[h!]
\centering
\includegraphics[scale=0.38]{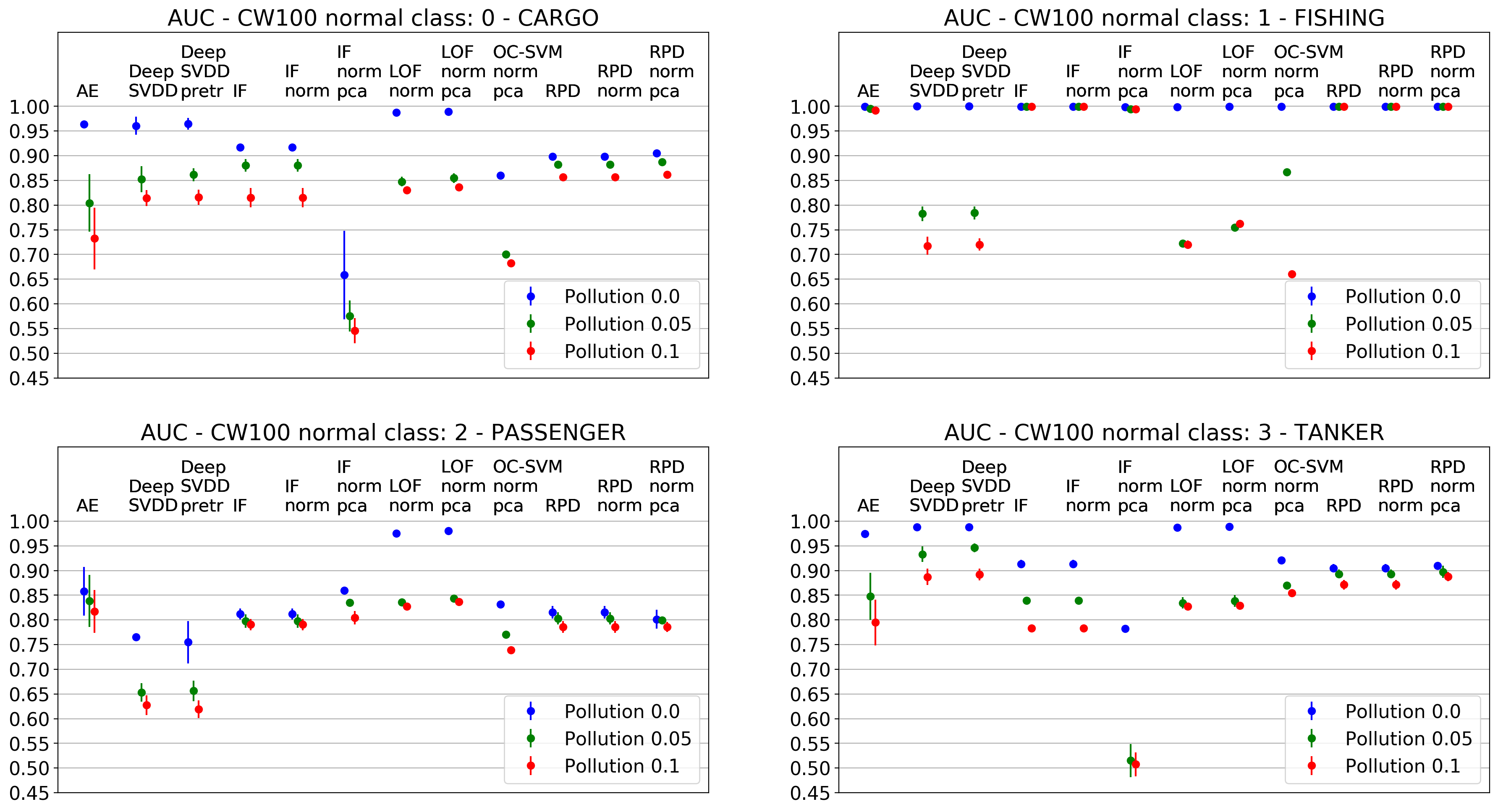}
\caption{Unsupervised AD results. Each color point describes an AUC averaged over the experiments where a single pollution class is considered among the anomalous classes at a time. For example, a point associated with the normal cargos will be defined by the mean AUC of the experiments where pollution anomalies stem from the passenger ships, the fishing ships and the tankers respectively. Also, each single experiment setup is executed on three different seeds. This implies that each point actually depicts a mean of mean (over varying pollution and seeds respectively). The vertical bar associated with each point represents the mean standard deviation over all experiments and seeds (one standard deviation is computed over each set of seeds, then a mean is determined over all experiments). Each method suffers a loss of AUC for a higher pollution ratio of its training set supposedly pure of anomalous samples.}
\label{fig_auc_unsupervised}
\end{figure*}


The results of unsupervised AD with training pollution are illustrated on Figure \ref{fig_auc_unsupervised}. Results are globally good, with an expected impact of progressive pollution of the supposedly purely normal training set. Weaker detection is achieved when the normal class consists of passenger ships. This class likely makes it harder to guess a normality boundary since it is associated with a great features variance, and ships belonging to a wider range of lengths. This interpretation is compatible with the smaller drop of AUC for this normal class when pollution is introduced in the training set: the anomaly detection was already a little confused by the variety within the normality before any pollution. This reminds us how essential it is to wisely choose what normality will be made of. The most sensible drops of performances can be seen for IF associated with normalization and PCA in preprocessing. A slight improvement can be observed when Deep SVDD benefits from a pretrained network thanks to an initialization that uses the weights of a CAE encoder trained on similar data. The most harmful pollution to normal classes cargo, passenger and tanker is the one made of fishing ships. Indeed, introducing anomalous fishing ships as pollution makes it harder to detect the most distinguishable class from normality in those cases.

IF and RPD stand out as the most stable AD through pollution (apart from IF with normalization and PCA), whereas the deep unsupervised AD methods and LOF indicate high sensibility to training set impurity, with an apparent plateau effect for LOF polluted experiments AUCs. LOF nonetheless obtains excellent results when the training set is not polluted. RPD also stands out as a shallow AD for which the normalization and PCA have no substantial impact on the AUC obtained. This stability illustrates the affine invariance properties of RPD. AUC is exceptionally high when the normal class is composed of fishing ships (class one) because of the easy distinction of this class among our three others: one could easily select most fishing ships in our dataset based on RP length only. RP size is perhaps the most reliable feature of our HRRPs since the latter are subject to high variability within a single class, the dominant scatterers changing between ships with the same function (\emph{i.e} the same AIS label), without even mentioning target-aspect sensitivity. Finally, regarding the resources needed to train the AD methods considered, this study revealed that outside the neural networks training that requires the biggest resources, LOF was the slowest, followed by OC-SVM (even though helped by a PCA-reduced dimensionality) and then IF. RPD was the quickest method to train in our study.

\subsection{Semi-supervised AD}

\begin{figure}[h!]
\centering
\includegraphics[scale=0.4]{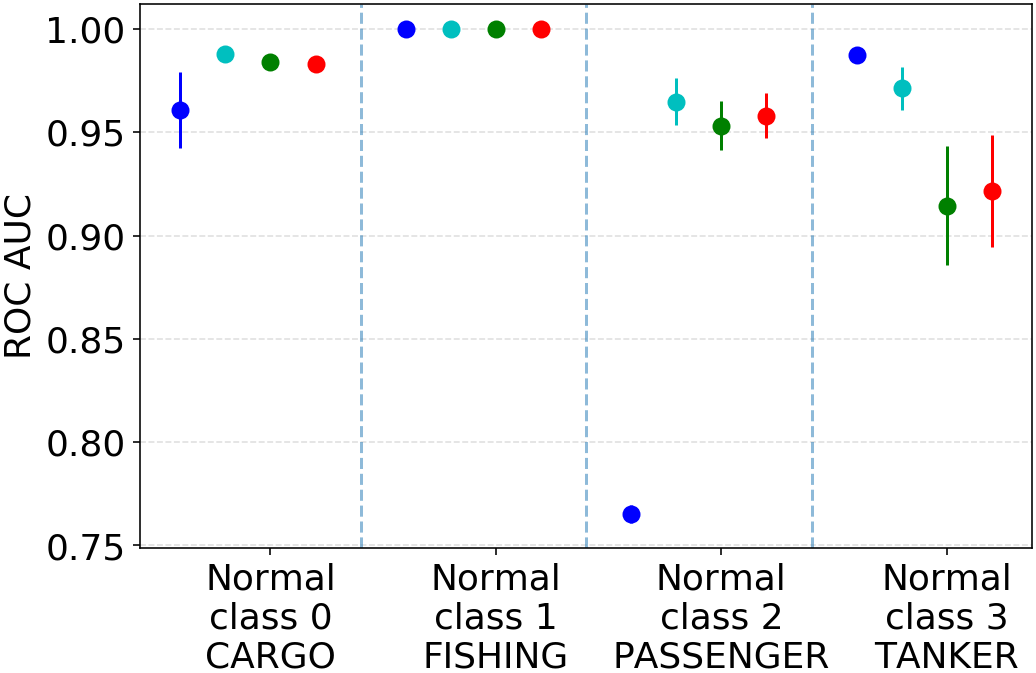}
\caption{Semi-supervised AD results using Deep SAD, the semi-supervised adaptation of Deep SVDD. AUC averaged over three experiments, with three seeds per experiment as for unsupervised AD: in each experiment we change the anomaly class from which labeled anomalies originate. Vertical bars represent standard deviations computed over the various experiments and seeds. The colors describe the four different labeled anomalies ratios in the training data considered (blue: 0\%, cyan: 1\%, green: 5\%, red: 10\%).}
\label{fig_auc_sad}
\end{figure}


The results of semi-supervised AD with labeled samples during training are illustrated on Figure \ref{fig_auc_sad}. For normal classes cargo and passenger, the introduction of labeled anomalies improves the mean AUC. One can further note that where unsupervised AD encountered difficulties for normal class passenger, SAD seems to tackle the latter. The mean AUC however drops with labeled anomalies for normal class tanker. Upon investigation, it emerged that a single SAD experiment (\emph{i.e.} a single type a labeled anomaly) is responsible for this drop: once labeled anomalies from class cargo are added to the training data when the normal class is tankers with a non zero ratio, the AUC drops. It seems not to stem only from excessive proximity of the two classes in terms of ship size otherwise the same drop would impact the AUC when the roles are reversed (normal class defined by cargo and contaminated by tanker). The reason behind this asymmetrical decrease could be a combination of the untreated HRRPs sensitivities and the intra-class ships diversity. Going back to the successful SAD experiments, the rise of AUC due to labeled anomalies for normal class cargo and normal class passenger ships appears to reach a plateau: 1\% and 5\% of labeled anomalies help approximately as much as 10\% labeled anomalies. A possible interpretation would be that few samples suffice to clarify the decision boundary of the AD method towards one specific kind of anomaly, with no further improvement in that so-called anomaly direction after a certain point. It is irrelevant to explore higher ratios of labeled anomalies within the training set, since doing so would switch our AD context with a supervised classification one.

\section{Conclusion}

This study has shown that a variety of anomaly detection methods can be effective for unusual HRRP targets detection. Our results on semi-supervised detection demonstrated the possibility to improve such detection using a few labeled anomalies. Besides, since hyperparameters were not extensively fine-tuned the methods could yield additional improvements. This potential is also increased by the various advantages and drawbacks offered by each method considered. An important feature of the methods presented is that they all produce a decision score that allows the radar operator to have a refined appreciation of the anomalous nature of an HRRP target, in contrast with a binary result. This echoes the call for more understandable machine learning, specifically regarding the confidence one can put in machine learning results. Future work will consider AD with multi-modal normality, scale-spaces generated invariance, and include other new and innovative, shallow and deep, anomaly detection methods such as memory augmented \cite{memae} and multiple-hypotheses \cite{multihyp} autoencoders. 

\section*{Acknowledgements}
%
This work was supported by the French Defense Innovation Agency (Cifre-Défense 001/2019/AID).

\bibliographystyle{unsrtnat}
\bibliography{mybibfile}

\end{document}